\documentclass[12pt]{article}

\usepackage{graphicx}
\usepackage{amssymb}
\usepackage{amsmath}
\usepackage{verbatim}
\usepackage{epsfig}
\usepackage{hyperref}
\usepackage{wasysym}
\usepackage{mathrsfs}

\newcommand{\beq}{\begin{equation}}
\newcommand{\eeq}{\end{equation}}
\newcommand{\beqa}{\begin{eqnarray}}
\newcommand{\eeqa}{\end{eqnarray}}

\newcommand{\cc}{{\rm c.c.}}
\newcommand{\mtridva}{{m_{\frac{3}{2}}}}
\newcommand{\xifi}{{\xi_{FI}}}

\begin{document}

\begin{titlepage}
\begin{flushright}
{\large UMD-PP-09-031}
\end{flushright}

\vskip 1.2cm

\begin{center}

{\LARGE\bf Breaking the Dark Force}

\vskip 1.4cm

{\bf \large  Andrey Katz$^a$ and Raman Sundrum$^b$}
\\
\vskip 0.4cm
{\it $^a$Department of Physics, University of Maryland,\\ 
College Park, MD, 20742}
\\
{\it $^b$Department of Physics and Astronomy, Johns Hopkins University,\\
Baltimore, MD, 21218}
\\

\vskip 4pt

\vskip 1.5cm

\begin{abstract}
\noindent  
 
Recently Arkani-Hamed, Finkbeiner, Slatyer and Weiner
 proposed a unified explanation 
to the set of experimental and observational anomalies with possible  
connections to Dark Matter (DM). A central role is played by 
GeV scale ``dark" gauge bosons exchanged by the weak scale DM.
Motivated by this proposal,
we build an explicit model of DM in the context of 
Weak Scale Supersymmetry (SUSY). We employ 
high-scale SUSY-breaking and invoke the Giudice-Masiero 
mechanism to generate the weak scale DM masses.
By sequestering the dark sector from the SUSY-breaking 
``hidden" sector, it naturally acquires GeV scale 
soft masses that help generate the dark gauge boson mass.
The visible MSSM sector is not fully sequestered 
and acquires gaugino-mediated soft terms at the weak scale.
This hierarchy of scales naturally leads to the Sommerfeld 
enhancement of DM annihilations needed to account for 
the $e^+ e^-$ excesses in the PAMELA 
and ATIC experiments.
The possibility of 
co-existing species of DM is used to show how the  
INTEGRAL and DAMA anomalies can both be explained. 
We study the cosmological constraints on the new 
stable or long-lived light particles that appear in 
our models. 
We discuss the lepton-jet collider signals suggested by 
Arkani-Hamed and Weiner, and find that 
they are not guaranteed
in our construction.

\end{abstract}

\end{center}

\vskip 1.0 cm

\end{titlepage}

\section{Introduction}
\label{sec:intro}
In recent years there have been an increasing number of experimental and 
observational anomalies which might possibly be due to the effects of
 Dark Matter (DM). INTEGRAL~\cite{Teegarden:2004ct} has detected a 511 keV 
emission line from the center of the galaxy due to positron annihilations, 
DAMA~\cite{Bernabei:2008yi} found annual modulations in their 
direct detection experiment, PAMELA~\cite{Adriani:2008zr} measured a positron 
excess in the high-energy  $e^+ e^-$ flux, while ATIC~\cite{Atic:2008zzr}
measured an excess in the overall $e^+ e^-$ flux at even higher energies. 
While any or all of these experiments might ultimately be explained 
without recourse to new physics, it is obviously of interest to 
tentatively consider them as clues  to 
the nature of DM, 
especially if such considerations motivate the search for 
new types of experimental signals.

If the DM explanation for DAMA and INTEGRAL  experiments are accepted, 
one should 
consider non-minimal DM scenarios. 
It is not easy to reconcile the DAMA measurements with other direct 
detection experiments. One of the notable explanations to 
this disagreement is the idea of inelastic dark matter 
(iDM)~\cite{TuckerSmith:2001hy}. On the other hand, the 511 keV line seen by
 INTEGRAL  can be successfully explained by the scenario of 
exciting dark matter (XDM)~\cite{Finkbeiner:2007kk,Pospelov:2007xh}. 
It is 
striking that these two phenomenologically distinct anomalies can 
be explained by similar mechanisms, namely roughly
${\cal O}({\rm MeV})$ splittings between DM states.\footnote{For
further discussion of some of the subtleties of these ideas 
see~\cite{Chen:2009dm,Finkbeiner:2009mi}} 

The explanation of PAMELA/ATIC signals via DM annihilation is also challenging: one requires DM annihilation cross section two to three orders of magnitude larger than needed to account for their measured relic 
abundance~\cite{Cholis:2008hb}.
Introducing  such large boost factors is somewhat implausible. 
Moreover, the absence of observable antiproton~\cite{Adriani:2008zq}
or gamma-ray excesses\footnote{The most stringent constraints on the DM models
explaining PAMELA are imposed by HESS
observations of the Galactic Center~\cite{Aharonian:2006wh} and the 
Galactic Ridge~\cite{Aharonian:2006au}. 
See~\cite{Meade:2009rb,Mardon:2009rc} for 
an exhaustive
discussion of the relevant constraints.}  
imposes stringent constraints on any 
explanation
in particular 
requiring suppression of hadron production from DM annihilations.

Recently a single attractive picture, 
explaining simultaneously all these phenomena, was suggested in~\cite{ArkaniHamed:2008qn, ArkaniHamed:2008qp}. 
The main idea is that the DM is charged under some intermediate-range gauge 
force with a gauge boson mass at the GeV scale, discussed earlier 
in~\cite{Pospelov:2007mp,Finkbeiner:2007kk}. 
This would result in a significant Sommerfeld enhancement of the 
late-time DM annihilation cross section, once DM is non-relativistic. 
 This framework also explains why 
there is an excess of positrons but no excess in the antiproton or photon 
fluxes. 
The main annihilation channel of DM is into the 
dark sector $\sim$ GeV-mass gauge bosons. 
The dominant interaction of the dark sector with the SM is naturally 
due to kinetic mixing
between the dark gauge boson and the regular photon (more fundamentally, 
hypercharge)~\cite{Holdom:1985ag}. 
If these gauge bosons do not have efficient decay channels within the dark 
sector, they will decay via this mixing to sub-GeV charged SM particles, 
leading mostly to
decays  to $e^{+} e^{-}$ and with few hard photons and no 
antiprotons.
The presence of such a light subsector interacting with DM can also 
explain the small splittings between the DM states, required for iDM and XDM. 
 These splittings can be either radiatively induced, or 
come from couplings between DM and the light fields. 

Inspired by the suggestion of~\cite{ArkaniHamed:2008qn} we develop in this paper a concrete model of the dark sector, within the paradigm of Weak Scale 
Supersymmetry.
Quite apart from the experimental anomalies discussed above, 
we are also 
motivated by trying to understand what simple connections 
might exist within SUSY field theory 
between our visible sector and other self-interacting ``dark'' sectors 
without SM-charges, in keeping with 
Hidden Valley ideas~\cite{Strassler:2006im}.

The central issue for the DM modelling 
is to find natural origins for the two scales 
of the dark sector, the DM mass scale of hundreds of GeV, and the GeV mass 
scale of the dark gauge boson. Clearly, we would like to connect both scales, 
as well as the weak scale, to the 
scale of SUSY breaking.
We follow closely the idea suggested in~\cite{ArkaniHamed:2008qn} that  DM
gets its mass from the same mechanism responsible for the appearance of 
the $\mu/B\mu$ terms in the MSSM, while the GeV scale emerges radiatively with
one-loop suppression, and MeV DM splitting emerge radiatively with 
two-loop suppression. 
 
What is new here is that 
we take the simplest, and in many ways the most attractive, approach to 
generating a $\mu$ term and DM mass, namely
(generalizations~\cite{Randall:1998uk} of) the 
Giudice-Masiero 
mechanism~\cite{Giudice:1988yz}.\footnote{For an earlier attempt 
to realize the model suggested
in~\cite{ArkaniHamed:2008qp} in the framework of gauge-mediation, 
see~\cite{Chun:2008by}.}   
This mechanism works best in the context of
high-scale SUSY-breaking. We will therefore consider the MSSM soft terms 
to be generated by gaugino-mediated SUSY 
breaking~\cite{Kaplan:1999ac,Chacko:1999mi},
 so as to resolve the 
SUSY flavor problem. But we are then faced with the question of why 
the dark gauge boson has mass of only GeV. If one introduces dark Higgs fields
to generate a mass, they should naively feel gravity-mediated soft masses of 
order $\mtridva \sim 100\ {\rm GeV}$, making a GeV scale Higgs mechanism 
quite unnatural.  
We resolve this puzzle in a simple way, by taking the entire dark sector to be 
sequestered from the SUSY-breaking hidden sector. 
Our set-up can therefore be pictured as in Fig.~\ref{fig:EDpict}.

The sequestered dark sector fields get their masses from several
different sources, but all of them are naturally suppressed compared 
to the scale $\mtridva$. First, the supersymmetric gauge
kinetic mixing also contains D-term mixing, resulting in an effective 
Fayet-Iliopoulos (FI) term for the dark 
sector after electroweak symmetry breaking~\cite{Baumgart:2009tn}. 
The strength of this mixing, $\epsilon$, can naturally be of one-loop to 
two-loop order if it is due to loops of heavy ``link fields'',
 carrying both dark charge 
and hypercharge (and possibly other SM gauge quantum numbers). 
With this suppression, the effective FI term in the dark 
sector is of order GeV$^2$. Furthermore, 
 the sequestered dark sector also gets  
anomaly-mediated (AMSB) soft masses~\cite{Randall:1998uk,Giudice:1998xp}. 
These masses are generically of order $\mtridva/(16\pi^2)\sim 1\ {\rm GeV}$. 
Moreover, IR-free gauge forces give tachyonic 
contributions to charged scalar masses in AMSB. 
These contributions combined with the effective FI term will 
break the dark force at $\sim$ GeV, as required.

Careful analysis of the dark sector in our construction reveals several  
interesting features, which seem quite generic.
We emphasize that almost all the models of this type
involve \emph{co-existing DM}. 
Besides the stable heavy DM (HDM), which 
is responsible for the experimental 
signals listed above, one finds a stable lightest R-odd 
particle at $\sim$ GeV (of course, if R-parity is assumed). 
This particle belongs to the dark sector and forms a light DM (LDM). 
In our model, the LDM is the lightest fermion of the dark sector,
which is thermally produced. This imposes a stringent constraint 
on the parameter space, to ensure that
the LDM has sufficiently strong annihilation channels,
such that its relic abundance is small.

The dark sector can also contain long-lived
particles. 
Indeed, if the dark gauge boson is not the lightest particle 
in the dark spectrum, the lightest particle 
decays induced purely by gauge kinetic mixing can be 
extremely slow. Such long-lived particles are severely constrained 
by Big Bang nucleosynthesis (BBN). 
However we identify simple new couplings between the dark and 
visible sectors that greatly speed up the decays. We also discuss 
how these couplings and the particle spectrum affect  
the ``lepton-jets" signatures 
predicted in~\cite{ArkaniHamed:2008qp}. 
We stress that these 
collider signals are not an inevitable outcome of our model. Dark particles produced at 
colliders can be sufficiently long-lived to escape the detectors before 
their decays to leptons.

Finally, in order to explain both DAMA and INTEGRAL
by iDM and XDM respectively,
we are required to introduce two flavors of  HDM. Because 
the HDM is charged only under an Abelian symmetry in our model, 
we do not have enough degeneracy in a single 
DM flavor to have both iDM and XDM splittings. While such co-existing HDM
is a plausible and interesting 
 possibility, we also note that our model might also be
extendable to accommodate non-Abelian dark gauge group, enabling a simultaneous
explanation for iDM an XDM from a single multiplet of DM.

The paper is organized as follows. In the Section 2, we assemble 
 our model in a modular fashion. 
In Section 3, we derive the spectrum of the dark sector taking into account 
SUSY breaking. In Section 4,  we discuss phenomenological 
requirements of our model and demonstrate
 viable regions in the parameter space. In particular we 
 explain how we get the required type of experimental 
signals for PAMELA/ATIC. 
We discuss the possible presence 
or absence of lepton jets in collider signals, and its sensitivity to 
the flavor structure of our new couplings between the dark and visible
sectors. Section 5 provides our conclusions.

While this paper was in preparation we learned about the work 
of~\cite{Princetongroup}, 
which has a considerable overlap with our work. While both our models 
consider a similar field content for the dark sector, the implementation 
of SUSY breaking is different.


\section{The Model}
\label{sec:model}

\begin{figure}[t]
\centering
\includegraphics{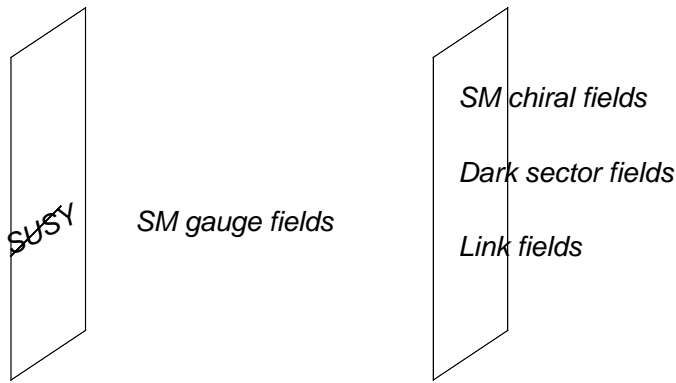}
\caption{A qualitative picture of extra-dimensional sequestering.}
\label{fig:EDpict}
\end{figure}

We consider high-scale supersymmetry breaking, gaugino-mediated to the 
visible sector, 
so that the mass scale of the SM superpartners is of order of the gravitino mass $\mtridva$. This requires sequestering of MSSM matter, but not gauge fields,
 from the SUSY breaking 
hidden sector. We also assume that the {\it entire} dark sector and any 
link fields are sequestered from the SUSY breaking hidden sector. 
The basic set-up is then shown in Figure~\ref{fig:EDpict}. 
 
Our dark sector is a simple generalization of SQED, with gauge 
group  $U(1)_D$. We choose some massive 
vector-like charged pairs, $X_i + \bar{X}_i$,  to house  
dark matter, and a massless vector-like charged pair, $T + \bar{T}$, to house the dark Higgs field responsible for breaking the $U(1)_D$ and
giving mass to the dark photon. For simplicity, all the charges are taken 
as $\pm 1$. We  also assign $\mathbb{Z}_2$ symmetries under which the 
$X_i, \bar{X}_i$  are odd, to ensure their stability (and non-mixing) when 
other interactions are added. All the chiral fields are taken R-parity even. 
None of them carry SM quantum numbers.

After $U(1)_D$ is broken, the safest option is to have the entire dark sector acquire mass. If there are massless particles in the dark sector, then dark matter 
annihilations will predominantly end up in these particles, and there is no 
chance of decaying  into the SM to give the PAMELA/ATIC signals. From this point 
of view, we can anticipate a general problem if we look at the fermions in 
$T, \bar{T}$. Without knowing the details of soft SUSY breaking in the
dark sector we can still conclude that one linear combination of these fermions will mix with the 
dark gaugino after the Higgs mechanism, giving mass to this combination in 
general. But the remaining linear combination of charged fermions remains massless, protected by the chiral symmetry of the massless fields. 
Of course we could add a small mass term for $T, \bar{T}$, but this 
would then require a tuned value, posing a kind of dark ``$\mu$-problem".
  Instead we will introduce a gauge singlet, $S$, whose fermion 
will marry the otherwise-massless charged fermion to form a Dirac state, 
once we include a superpotential term
\beq\label{hiddenW}
W =\lambda S T \bar T + \frac{\kappa}{3!} S^3.
\eeq
Of course $S$ is also taken as even under all discrete symmetries.

We consider the dark gauge coupling, $g_D$, as well as $\lambda, \kappa$ 
to be roughly order one.

\subsection{\it Dark Matter Mass}

We do not put in  direct superpotential mass terms in the dark sector, 
because just as for the MSSM Higgs multiplet, this would lead to a 
$\mu$ problem, or coincidence of scales arising from disparate sources.
Of course, since such mass terms are not renormalized and are protected by 
chiral symmetries, their vanishing below the Planck scale is a natural 
possibility.

We are assuming that the entire dark sector is sequestered from the 
hidden sector responsible for intermediate scale supersymmetry breaking, 
so we use a simple generalization of the Giudice-Masiero 
mechanism~\cite{Randall:1998uk} to 
give weak-scale masses to the $X$'s. Their Kahler potential is given 
in off-shell SUGRA by
\beq
 K =|\phi|^2\left( |X|^2+|\bar X|^2 +c_i (X_i\bar X_i+\cc) \right),
\eeq 
where the VEV of the SUGRA  conformal compensator is given by
\beq
\phi\equiv 1+\mtridva \theta^2.
\eeq 
We take the natural size of dimensionless coefficients $c_i \sim {\cal O}(1)$.
One can think of them as arising at the Planck scale in the form $ {\cal O}(1) 
\langle \chi^{\dagger} \rangle/M_{Pl}$, where $\chi$ is a chiral field with 
$\sim$ Planckian mass and VEV that breaks the chiral symmetries that protect 
a direct $X$ mass.

After superfield rescaling $X\phi\to \phi$ and $\bar X \phi\to \phi $ we get the effective K\"ahler potential 
\beq\label{DMKahl}
K = |X|^2 +|\bar X|^2 + c \frac{\phi^\dagger}{\phi} (X\bar X+\cc)
\eeq
resulting in $\mu$ and $B\mu$ terms 
\beq
\mu_i=c_i \mtridva,\ \ \ B_i \mu_i =c_i\mtridva^2.
\eeq
Assuming that the MSSM acquires SUSY breaking through gravity-mediation or
gaugino-mediation implies that $\mtridva$, and therefore $\mu_i, B_i$, are of order the weak scale. Note that these ``soft" masses have emerged despite 
sequestering of the dark fields (no direct coupling to the hidden sector).

The tree-level spectrum of $X + \bar{X}$ is given by
\begin{eqnarray}
m_{1/2} &=& c \mtridva \nonumber \\
m_{0, X + \bar{X}^\dagger}^2 &=& c(c + 1) \mtridva^2 \nonumber \\
m_{0, X - \bar{X}^\dagger}^2 &=& c(c - 1) \mtridva^2.
\end{eqnarray}
We take $|c_i| > 1$ to avoid scalar tachyons. We take the 
analogous $c$ coefficient to vanish for our dark Higgs sector, $T + \bar{T}$, to 
avoid getting mass. Again, this is a natural possibility.

\subsection{\it Mixing with SM Hypercharge}

The renormalizable kinetic mixing of $U(1)_D$ and hypercharge is given by
\beq\label{kinetmix}
{\cal L} \supset \frac{\epsilon}{2} \int d^2 \theta W^\alpha_{D} W_{\alpha Y}+\cc.
\eeq
We assume that $\epsilon$ is induced by loops of heavy ``link fields"
carrying both SM and dark charges, with a natural 
range of $\epsilon \sim 10^{-4}\ - \ 10^{-3}$. 

For the most part such a tiny mixing has little effect on the dark sector, 
but there are three qualitatively important exceptions which we can anticipate: 

(i) As suggested by~\cite{ArkaniHamed:2008qn} otherwise stable particles in the dark sector may decay into the SM via this tiny mixing. This is the key to the PAMELA/ATIC signals from dark matter annihilations in their proposal. 

(ii) The mixing allows photon-dark-photon exchanges to mediate direct 
detection of dark matter. The smallness of the mixing will be offset by 
the lightness of the mediating particle relative to say $Z$-exchange in 
more standard scenarios. 

(iii) As noted in~\cite{Baumgart:2009tn}, the $D$-terms of the  $U(1)$'s are also mixed, 
and electroweak symmetry breaking in the SM leads to an effective Fayet-Iliopoulos $D$-term in the dark sector:
\beq\label{effi}
\xifi = \epsilon D_Y,\ \ D_Y=\frac{g' v^2 \cos (2\beta)}{2}
\eeq
where $D_Y$ denotes the D-term of the SM hypercharge and $v$ is the electroweak scale $176\ {\rm GeV}$. 
For the range of $\epsilon$ considered this introduces the 
desired GeV scale into the dark sector. 

Note that the decays to the SM via (i) are most efficient if the 
dark photon is stable for $\epsilon = 0$, since it is 
directly coupled to the SM photon by the mixing. In the present paper
we will consider this to be a subdominant channel for decays to the SM,
to satisfy phenomenological constraints discussed in Section~4. 
In the next subsection we will 
introduce other simple couplings between the two sectors which will dominate 
the PAMELA/ATIC signals from dark matter annihilations. 
The kinetic mixing $\epsilon$  will however continue to 
serve purposes (ii) and (iii). 

\subsection{\it  Singlet couplings to the SM }

We will be led to 
study a part of the parameter space in which the singlet scalar 
is the lightest particle, into which other dark gauge/Higgs degrees of
freedom rapidly decay or annihilate. The PAMELA/ATIC signals can then 
arise from decays of the singlet scalar into the SM. Rapid decays 
can arise from non-renormalizable direct couplings of the form
\beq\label{yukawamix}
{\cal L} \supset \int d^2 \theta \left( \frac{S L   H_d \bar e}{\Lambda} + 
\frac{S \bar d   H_d Q}{\Lambda}\right),
\eeq
where $L, \bar e, Q, \bar d$ denote the usual 
weak doublets and down-type singlets 
of the MSSM and $H_d$ is the down-type Higgs.  We might
expect the coefficients of these couplings to be roughly 
similar to Yukawa couplings,
while the overall suppression scale $\Lambda$ is due to integrating out some 
high-scale physics. 

A simple example of such high-scale physics is a vector-like pair of
chiral multiplets carrying SM quantum numbers with Yukawa couplings to the 
SM and S. To preserve perturbative unification we can take the pair to
consist of a massive lepton doublet\footnote{Somewhat similar idea of 
introducing new leptons was suggested in~\cite{Phalen:2009xw}. Note that in 
our case these fields can be much heaver than $\sim 100\ {\rm GeV}$ as 
was proposed in~\cite{Phalen:2009xw}.},  
${\mathscr L} + \bar {\mathscr L}$,
 and massive down-type quark singlet, $ {\mathscr D} +\bar {\mathscr D}$,
forming a $5+ \bar 5$ of $SU(5)$. Given superpotential couplings of the form
\beq\label{rensinglcoupl}
W = SL \bar {\mathscr L}  + \bar {\mathscr L} H_d \bar e + S\bar d {\mathscr D}
+ \bar {\mathscr D} H_d Q +M_{{\mathscr L}} {\mathscr L} \bar {\mathscr L} +
M_{{\mathscr D}} {\mathscr D}\bar {\mathscr D},
\eeq     
we get~\eqref{yukawamix} upon integrating out the mass 
$M_{{\mathscr L}}, M_{{\mathscr D}}$ states.

Upon electroweak symmetry breaking, the effective couplings~\eqref{yukawamix} 
give the scalar $S$ decay channels into kinematically available pairs of 
 SM leptons and quarks. There is a large range of possible 
masses $M$ or scales $\Lambda$ over which 
these decays can be very rapid. We will return to discuss the constraints on 
these decays and their phenomenology in Section~4.

\subsection{\it Dark Matter splittings for iDM and/or XDM}

Our stable dark matter particles are the complex dark-charged scalars 
$X_i - \bar{X}^{\dagger}_i$. For iDM or XDM to work we need to split the 
real and imaginary components of these scalars which are 
connected by dark-photon couplings, by of order MeV. 
If we want {\it both} iDM and XDM we will need sub-MeV and 
super-MeV splittings, and therefore we will simply invoke 
two co-existent dark matter species, $i = 1,2$. If only one or 
the other of iDM or XDM is needed, one species of dark matter, 
$i=1$, will suffice. 

Real-imaginary splittings can emerge once the dark matter 
feels the breaking of $U(1)_D$, say via the coupling to 
the dark Higgs, 
\beq\label{nonreni}
{\cal L} \supset {\cal O}(1) \frac{\int d^2 \theta  T^2 \bar{X}^2}{\rm TeV}
\eeq
Such non-renormalizable couplings can readily arise 
by integrating out  $ \mathbb{Z}_2$-odd singlets, $A_i$, 
with weak scale masses via the (conformal compensator version of the) 
Giudice-Masiero mechanism, and couplings of the form
\beq\label{amix}
{\cal L} \supset \eta \int d^2 \theta  ~ T \bar{X}_i A_i~, 
\eeq
where we define a coefficient $\eta$ for later reference.

Now we have introduced a full theory of the dark sector. Adding the 
(gauged) kinetic terms for all the fields, together with the equations~\eqref{hiddenW},~\eqref{DMKahl},~\eqref{kinetmix},~\eqref{yukawamix},~\eqref{nonreni} one gets the full effective 
 Lagrangian of the dark sector and its interactions with the visible sector, 
to be used in the analysis below.    


\section{Dark spectrum}
In this section we will determine the ground state of the dark sector in various parts of the parameter space as a consequence of the SUSY breaking. We will then study higgsing of $U(1)_D$ and the mass spectrum.      

\subsection{\it SUSY breaking contributions to the dark sector}
SUSY breaking enters the dark sector from three different sources:
\begin{itemize}
\item Effective Fayet-Iliopoulos term
\item standard AMSB
\item deflection from AMSB due to non-decoupling threshold effects 
\end{itemize}
Below we discuss each of these contributions.

As discussed in the previous section gauge kinetic mixings, electro-weak and SUSY breaking in the visible sector induce an effective FI term
\beq\label{fiterm}
{\cal L}\ \supset \xifi \int d^4 \theta V
\eeq 
for the dark sector~\eqref{effi}.    
As we have already mentioned for the values of $\epsilon$ of order $10^{-3} \ - \ 10^{-4}$,  $\xifi\sim (1\ {\rm GeV})^2$. This contribution is the only 
non-sequestered one in the dark sector. (Sequestering would be 
exact in the $\epsilon \rightarrow 0$ limit.) 
 While it is triggered by SUSY breaking effects in the visible sector, from the dark sector point of view this contribution is entirely supersymmetric~\eqref{fiterm}
 
Because of sequestering the remaining SUSY breaking contributions must arise from AMSB~\cite{Randall:1998uk,Giudice:1998xp}. We are interested primarily in 
soft terms for the light degrees of freedom. The usual form of such contributions depend only on the IR effective theory and $\mtridva$.  
 These contributions are of order one-loop for the gaugino masses and A-terms and of order two-loops for the scalar masses squared, corresponding to the ${\cal O}(1\ {\rm GeV})$ scale again, given $\mtridva \sim {\cal O}({\rm 100 GeV})$.

In AMSB, heavy thresholds due to a supersymmetric mass decouple from IR soft terms. Naively, such  decoupling would apply to the HDM contributions to IR soft terms. However this is not true for two reasons. Firstly, 
such decoupling is only true when the thresholds are much larger than $\mtridva$, while our HDM is comparable. Secondly, mass thresholds 
due to mass terms in a superpotential satisfy 
\beq
\frac{B\mu }{\mu} = \mtridva,
\eeq
while mass thresholds from the Kahler potential such as~\eqref{DMKahl} have a sign difference
\beq
\frac{B\mu}{\mu} = {\bf -} \mtridva.
\eeq  
While this sign difference is important for gauginos, it cancels in the 
dominant scalar soft terms~\cite{Nelson:2002sa}.

\subsection{\it Mass spectrum}
Now we are ready to write down the scalar potential of the light dark sector. A \emph{supersymmetric} D- and F-term scalar potential, including the effective FI term, reads
\beq\label{hiddensp}
V=\frac{g_D^2}{8}\left( |T|^2 - |\bar T|^2 - \frac{\xifi}{g } \right)^2+\left|\frac{\kappa}{2}S^2+\lambda T \bar T \right|+\lambda^2\left( |TS|^2+|\bar T S|^2\right)~. 
\eeq

Soft masses  are 
\beqa\label{softamsb}
m_T^2 & = & \frac{\mtridva^2}{(16\pi^2)^2} \left( -4 g_D^4+3|\lambda|^4 -4  |\lambda|^2 g_D^2 +\frac{|\lambda|^2|\kappa|^2}{2}\right)- \frac{1}{6} \frac{\eta^2}{16\pi^2}\left( \frac{\mtridva^2}{c^2} + \frac{\mtridva}{c_A^2} \right)\\
m_S^2 & = & \frac{\mtridva^2}{(16\pi^2)^2} \left( 3|\lambda|^4 -4 |\lambda|^2 g_D^2+2 |\lambda|^2 |\kappa|^2+\frac{3}{4}|\kappa|^4 \right)~. 
\eeqa  
An A-term which couples the scalars $T,\bar T,\ S$ is also an important contribution to the spectrum because $T$ condenses: 
\beq\label{aterm}
a\equiv a^{T\bar TS} = -\frac{\mtridva}{16\pi^2}\left( 3\lambda |\lambda|^2 - 4\lambda  g_D^2 +\frac{\lambda |\kappa|^2}{2} \right)
\eeq   
The first term in the expression~\eqref{softamsb} is a standard AMSB contribution while the second one comes from Yukawa couplings 
to HDM~\eqref{amix} 
with non-decoupling masses~\cite{Katz:1999uw}.   
We also get soft gaugino mass, which reads (including the non-decoupling threshold~\eqref{DMKahl})
\beq
m_{1/2}=\frac{g_D^2 \mtridva (N-1)}{8\pi^2}~.
\eeq 
Here $N$ denoted a number of charged HDM flavors. 

First, in order to simplify the following analysis we consider $c\sim 10$ so that the non-decoupling Yukawa mediated contribution ($\eta^2$ contribution to 
soft masses)
is subdominant. Clearly for $\lambda \gg g_D$ neither $T$ nor $\bar T$ condense, so we take $\lambda<g_D$. The masses from AMSB are the same for $T$ and $\bar T$, but the contributions of the FI term have opposite signs. For simplicity we will restrict ourselves to the regime, where only $T$ condenses. We stress that there is no strong physical reason for restricting to this regime. Indeed even for $\epsilon=0$ and hence vanishing FI term, we can achieve viable $U(1)_D$ breaking when all three fields $T,\ \bar T,\ S$ get VEVs.

If only one field gets a VEV, its precise value is
\beq
\langle T \rangle = \sqrt{\frac{\xifi}{g_D}- \frac{4 m_T^2}{g_D^2}}~.
\eeq
\begin{figure}[t]
\centering
\includegraphics[width=0.8\textwidth]{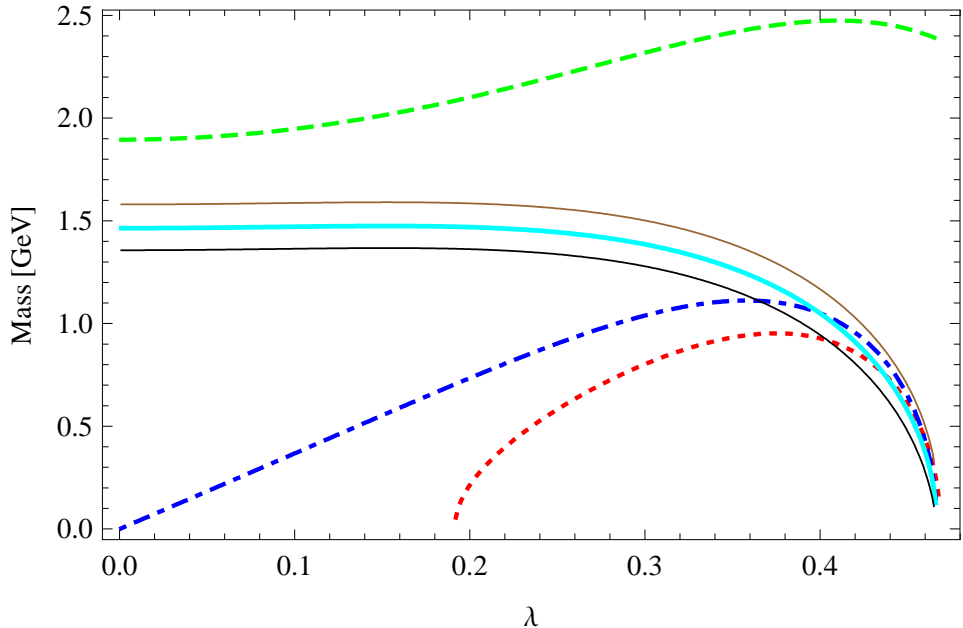}
\caption{The mass spectrum of the dark sector, measured in units of GeV, plotted against various values of $\lambda$. The values which have been chosen for this plots are: $g_D=0.4,\ \mtridva=110\ {\rm GeV}, \kappa=1,\ \epsilon=10^{-4},\ \tan \beta= 10$. Here, as well as on the following plots, the red (dotted) and the green (dashed) lines denote the scalars from $S$ and $\bar T$ 
admixtures, the blue (dashed-dotted) line denotes a Dirac fermion from $S$ and $\bar T$.  A cyan solid thick line denotes the mass of the gauge boson, while the brown and the black thin lines denote Majorana fermions from the gauge 
multiplet and $T$.}
\label{fig:spect1}
\end{figure}
\begin{figure}[t]
\centering
\includegraphics[width=0.45\textwidth]{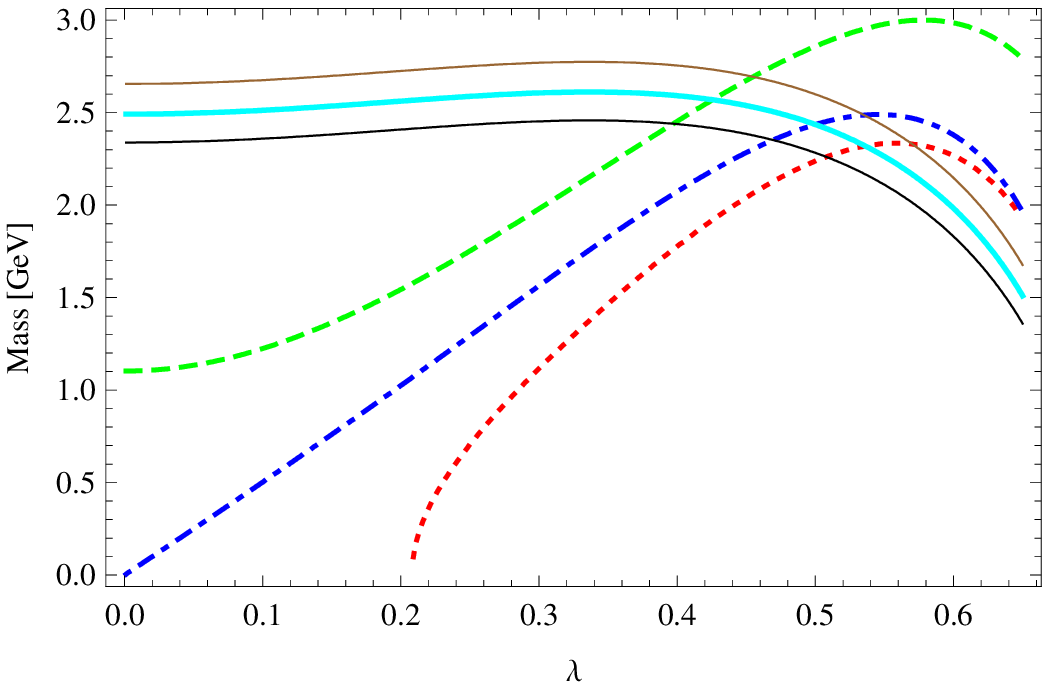}
\hspace{0.3cm}
\includegraphics[width=0.45\textwidth]{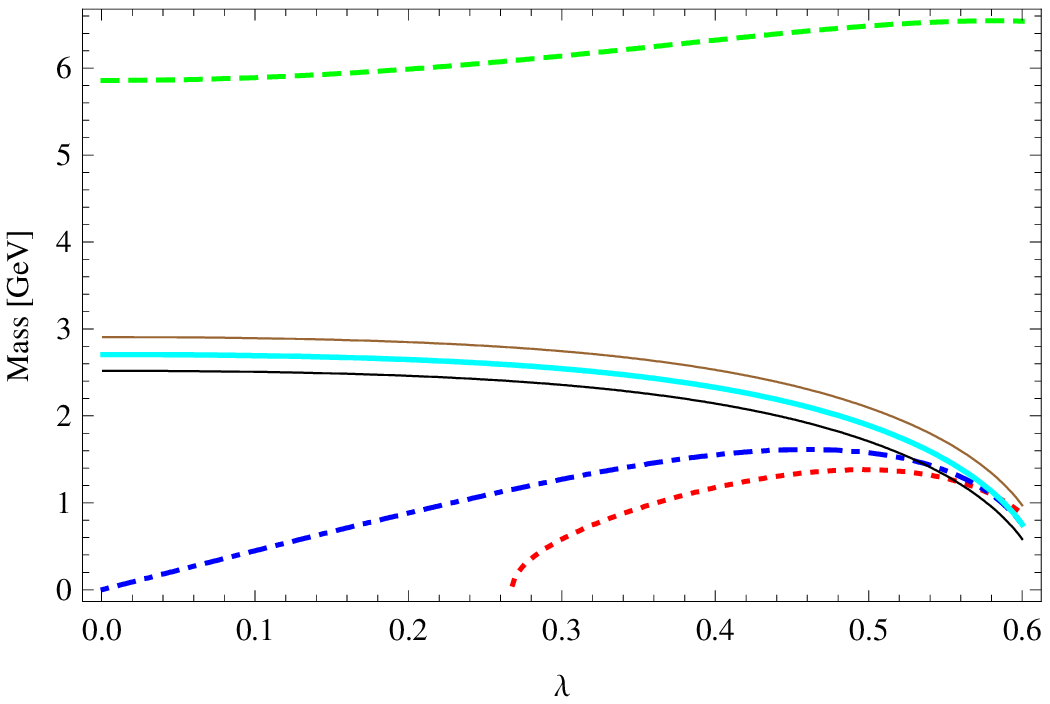}
\caption{The mass spectrum of the dark sector. The values 
 chosen for the right-hand side picture are  $g_D=0.5,\ \mtridva=100\ {\rm GeV},\ \kappa=0.8,\ \epsilon=10^{-3},\ \tan \beta=8$ and for the left-hand side picture $g_D=0.6,\ \mtridva=85\ {\rm GeV},\ \kappa=2,\ \epsilon=5\times 10^{-4},\ \tan \beta=30$. Again the mass is measured in units of GeV, the red
dotted line and the green dashed line denote scalars from $S$ and $\bar T$
admixtures, the blue dased-dotted line denotes a Dirac fermion from $S$ and $\bar T$. A cyan solid thick line denotes the mass of the gauge boson, while the brown and the black thin lines denote Majorana fermions from the gauge multiplet and 
$T$.}
\label{fig:spect2}
\end{figure}
Now we turn to the spectrum of the dark sector. Let us first analyze the masses of the fields in the gauge supermultiplet. When $U(1)_D$ is broken, the gauge field eats the phase of  $T$. The mass of the gauge field is 
\beq
m = g_D \langle T \rangle~.
\eeq
If the sector had been supersymmetric, this also would be a mass of the fermions.  
The photino mixes only with the $T$-fermion after $T$-condensation and one can write down $2\times 2$ mass matrix for these fermions:
\beq
m = \left(\begin{array}{cc}
m_{1/2} & g_D\langle T \rangle \\
g_D\langle T \rangle  & 0
\end{array} \right)
\eeq  
Diagonalizing this matrix we find two Majorana states with the masses
\beq\label{majmass}
m=\frac{1}{2} \left| m_{1/2}\pm \sqrt{g_D^2 |T|^2 +m_{1/2}} \right|~. 
\eeq
The fermion from the $S$ multiplet joins the fermion from the $\bar T$ multiplet in order to form a Dirac state with the mass $\lambda T$. 

The scalars of $S$ and $\bar T$ mix with each other by virtue of the A-term, but they do not mix with the physical scalar $T$. Therefore their masses squared are given by the eigenvalues of the following  $2\times 2$ complex matrix
\beq
m^2 = \left( \begin{array}{cc}
|\lambda|^2 |T|^2 +m_T^2 & a T\\
a^* T^* & |\lambda|^2 |T|^2+m_S^2 
\end{array} \right)
\eeq
The eigenvalues of this matrix are 
\beq
m^2 = \frac{1}{2}\left(2 |\lambda|^2 |T|^2+m_S^2+m_T^2\pm \sqrt{(m_T^2-m_S^2)^2+ 4 |a|^2 |T|^2} \right)
\eeq 
Mostly in the parameter space, both these mass-squared are positive. Rendering one of these masses tachyonic signals the breakdown of the single-VEV solution.  

We plot the spectra for various choices of parameters in figures~\ref{fig:spect1} and~\ref{fig:spect2}. Note that for each choice of the parameters, for sufficiently small $\lambda$ the lightest scalar becomes massless and below this value becomes tachyonic, signalling the onset of multi-VEV solutions, which we do not consider. 
In the vast part of the parameter space a scalar admixture of $S$ and $\bar T$ is the lightest particle, although there is a region in which the lightest particle is one of the Majorana fermions. As we will see in the next chapter, the first possibility is much more preferable.  
 

\section{Phenomenology}
\label{sec:experiment}

Below we list the phenomenological issues that constrain our parameter space.

\begin{itemize}

\item Dark matter abundance. We should ensure that  the correct dark matter 
density of the Universe can be accounted for as the thermal relic 
abundance of HDM.

\item PAMELA/ATIC signal. We should provide an effective leptonic decay channel for one of the light particles in the dark sector. If the gauge boson is not the lightest particle in the spectrum, this feature is not, a-priori, guaranteed. 

\item LDM relic abundance. As we have pointed out, at least one state in the dark sector is absolutely stable since it is the lightest R-odd particle. One should worry that this relic is at least not overproduced. Ideally we would like to render the abundance of this relic negligible. 
 
\item Constraints from BBN. In some parts of the parameter space the lightest \emph{unstable} particle is not the dark photon. If this particle is too long-lived, one finds severe constraints from the observed light elements 
abundance~\cite{Cyburt:2002uv}. We should verify that our model is not excluded due to these constraints.  

\end{itemize}

\subsection{\it Who should be the lightest particle?}
Observing the spectra of the dark sector (see, for example, figure~\ref{fig:spect2}) we distinguish between two qualitatively different cases. For relatively small $\lambda$ we find that a scalar admixture of $S$ and $\bar T$ is the lightest particle. The next to lightest particle is a Dirac fermion. When $\lambda$ increases one of the Majorana fermions becomes the lightest state and the gauge boson is the next-to lightest state. 

The lightest fermion of the hidden sector is always stable, because it is the lightest R-odd particle. We must verify that the abundance of this particle in the Early Universe is negligible. When scalars are the lightest dark particles
 there is no problem with the relic fermion abundance.
 The annihilation cross section of fermions to scalars is of order:
\beq
\langle \sigma v\rangle \sim \frac{\alpha_\lambda^2 
|\lambda \langle T \rangle |^2}{m_T^4} 
\eeq  
It is clear that this cross section is large enough for the fermions to 
annihilate away to scalars, and we will ensure that the scalars decay 
rapidly. Therefore,
 we do not have any problems with the LDM overproduction. 

But if the Majorana fermion is the lightest particle we have  a problem to wash it out efficiently enough. The only kinematically allowed 
annihilation channels are into the SM particles, which 
 are strongly suppressed by $\epsilon$, and
the annihilations into the leptons are also p-wave suppressed.

We conclude that the small $\lambda$ region is phenomenologically safest,
where scalars are the lightest dark particles. 
  
\subsection{\it Decays into the SM: astrophysical implications}

We will call light dark particles that cannot decay within the dark sector, ``dark-stable". Such particles are either absolutely stable, forming LDM, 
or can decay to the SM via the very weak  dark-visible 
couplings~\eqref{kinetmix},~\eqref{yukawamix}. 
They are important for two reasons. Given that HDM can annihilate 
into any of the dark-stable particles, the dark-stable decays to the 
SM can account for a variety of astrophysical signals. At colliders, 
R-parity ensures that sparticle production ends in decays into the 
LDM, potentially with associated production of other dark particles. 
If these latter particles are dark-stable they can decay back to the SM, 
possibly within the detector. 

The dark photon is dark-stable in 
some parts of parameter space where it cannot kinematically decay into lighter 
particles, as can be seen in fig.~\ref{fig:spect1}. In this case, it 
nevertheless promptly decays to the SM via gauge kinetic mixing, as discussed in~\cite{ArkaniHamed:2008qp}. However, note that the dark photon can 
easily be heavier than a GeV, which allows it to have both leptonic and 
hadronic decay modes (eg. $\omega^0+\pi^0$). As  emphasized 
in~\cite{Meade:2009rb}, if this channel accounts for the entire PAMELA/ATIC 
lepton signals, then it should also produce enough photons from hadron decays
to be in conflict with HESS measurements. But in our model, this is not the 
primary source of the PAMELA/ATIC leptons.

Indeed, the lightest dark particle in our chosen parameter region is a 
dark scalar, which is automatically dark-stable. Most of the HDM annihilation 
channels contain it as one of the final dark states. It can then decay into the 
SM and account for most of the astrophysical signals. If such decays 
are predominantly to $\mu^+ \mu^-, e^+ e^-$ and sufficiently 
dominate over the dark photon decays from HDM annihilation, then the 
photon bounds from HESS can be satisfied. More detailed study is needed 
to determine the viable parameter space. An obviously safe parameter region 
is where $m_{\gamma_D} > 2 m_0$, so that the dark photon is {\it not} 
dark-stable.  

Let us now turn to this lightest particle of the dark spectrum, a scalar. We should verify that it decays efficiently enough in order to produce a signal for PAMELA and ATIC and that it decays fast enough to evade the BBN bounds.  
Here we will explicitly show that $\epsilon$ cannot do the 
job by itself and we indeed need other couplings between the dark and visible 
sectors, such as the effective operator~\eqref{yukawamix} discussed in
Section 2.

\begin{figure}[t]
\centering
\includegraphics{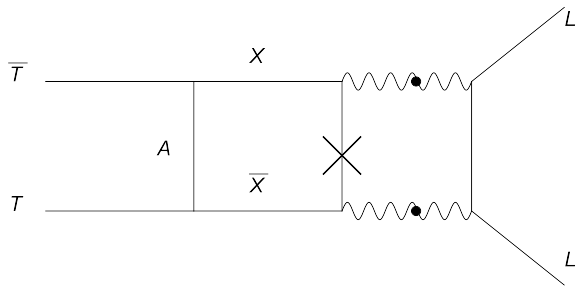}
\caption{An effective operator, responsible for the decay of $\bar T$
 is formed at two-loop level (supergraph notations has been used). The insertions on the photonic lines denote $\epsilon$ and the cross denotes $\mu$-term insertion.}
\label{fig:ttll2oop}
\end{figure}

To see this consider first the theory where $\epsilon$ alone connects the 
visible and dark sectors. The dark superpotential is given by
\beq\label{ma}
W_{eff} = \lambda T\bar T S +\frac{\kappa}{3!}S^3+ \eta TA\bar X +\bar \eta \bar T AX+\lambda T\bar T S +\mu_A A^2 +\mu X \bar X
\eeq    
The light fields $T,\ \bar T,\ S$ by themselves couple with an accidental 
$\mathbb{Z}_3$ symmetry under which they transform as
\beq
\bar T \to e^{\frac{2\pi i}{3}} \bar T,\ S \to e^{-\frac{2\pi i}{3}}S,\ T\to T~.
\eeq
This symmetry survives $T$ condensation and forbids the decay of the light 
scalar made from $S, \bar{T}$. The decay therefore requires an HDM loop 
since it is couplings to these heavy states that break the symmetry. 
The dominant decay process is then given by Fig.~\ref{fig:ttll2oop},
of course when the external leg of $T$ is replaced by its VEV. 
(One cannot directly decay 
to SM massless photons since the $\epsilon$ vertex is proportional to 
$p_{photon line}^2$ which would vanish for on-shell photons.) A rough estimate 
yields a lifetime of order  
 $\tau\sim 1000\ {\rm years}$. 

A particle with such a long lifetime will decay into the charged SM particles 
long  after BBN is over. Such decays would trigger EM showers in the plasma of the Early Universe and cause new processes of creating and destroying light elements. As has been shown in~\cite{Cyburt:2002uv} the decays of our light scalar will cause severe overproduction of deuterium. 

This problem can easily be  circumvented 
when there are new, even very weak, couplings between visible and dark
sectors, such as ~\eqref{yukawamix}. For a large range, 
$\Lambda< 10^{11}\ {\rm GeV}$,
assuming such couplings have coefficients which are roughly the same 
as the corresponding Yukawa couplings, a scalar $S \apprle 600$ MeV  would 
decay to a muon pair in less than a second, before the start of BBN.\footnote{Analysing more carefully the constraints from BBN we notice that this demand can be even further relaxed. The authors of~\cite{Cyburt:2002uv} 
impose strong bounds due to the late photo-dissociation
processes only for particles with  lifetime $\tau \apprge 10^4\ {\rm sec}$. Another constraint, which comes from 
entropy production considerations, does not apply in our case due to relatively small number density of the unstable relic. This means, that the scale $\Lambda$ can be even as high as $10^{15}\ {\rm GeV}$.} 
The muons in turn 
would decay to the electrons and positrons of the PAMELA/ATIC signals.

Of course, we must also estimate unwanted hadronic decays of $S$.
Antiprotons are kinematically forbidden for such light $S$, but  decays to
 $\pi^0$'s would produce photons. Such decays would proceed via the coupling
to strange quarks, $S \bar s s$,  comparable to the 
$S \bar{\mu} \mu$ coupling if there is a rough proportionality to 
SM Yukawa couplings. At leading order in chiral perturbation theory 
$m_s \bar s s \equiv m_{K}^2 |K|^2$, but kaon pairs 
are kinematically disallowed in $S$ decays. Pairs of $\pi^0$ are not 
interpolated by $\bar s s$ 
at leading order in chiral perturbation theory but they do appear at 
the next order in the derivative expansion, 
which by naive dimensional analysis yields 
\beq
m_s \bar s s \sim \frac{m_K^2}{(4 \pi f_{\pi})^2 f_{\pi}^2} 
(\partial_{\nu} \pi^0)^2.
\eeq
(The non-derivative coupling of the pions is subdominant if $S$ is heavier 
than the pion pair.)
We then arrive at the branching ratio estimate
\beq\label{decratio}
\frac{\Gamma_{S \rightarrow \pi^0 \pi^0}}{\Gamma_{S \rightarrow \bar{\mu} \mu}}
\sim \frac{1}{16} \left( \frac{M_{\mathscr L}}{M_{\mathscr D}} \right)^2
\left( \frac{m_s}{m_\mu}\right)^2.
\eeq
The factor of 16 arises from the enhancement of the muon channel due to 
decay into distinct Dirac fermions rather than identical real scalars,
and the derivative coupling of the scalars. The rough proportionality  
in the 
operators~\eqref{rensinglcoupl} to the SM Yukawa couplings leads to the 
appearance of the lepton to quark mass ratio in~\eqref{decratio}.
The $M_{{\mathscr L}}/M_{{\mathscr D}}$ ratio accounts for the possibility
of different suppression scales in equtation~\eqref{yukawamix} for 
quarks and leptons. 

We should bear in mind that this is just a rough estimate, depending on our
assumption
that the couplings $\eqref{yukawamix}$ are proportional to 
SM Yukawa couplings. 
But in any case the free parameter $M_{{\mathscr L}}/M_{{\mathscr D}}$
can safely suppress this branching ratio. 
This provides one
simple explanation for the 
excesses observed in PAMELA and ATIC, without attendant excesses of photons 
or antiprotons. 

Of course, for scalar $S$ mass below about $300$ MeV, the $\pi^0 \pi^0$ 
is suppressed simply by phase space considerations, and 
for $S$ mass below $200$ MeV  $e^+ e^-$
channel strongly dominates. 
These are also viable possibilities. But as one can see from 
Figs.~\ref{fig:spect1},~\ref{fig:spect2}
they arise in rather small parts of our parameter space. 
It may however be more readily occurring in the part of the parameter 
space in which there are more dark VEVs than just $\langle T \rangle$,
although we have not done this analysis.

Finally, we mention that the heavier dark fermions (R-odd) can also be 
kinematically dark-stable, and therefore can also decay to the SM. 
In particular, they can decay to the LDM and a lepton pair via 
a virtual exchange of the lightest dark scalar. Its lifetime is only 
an inverse loop factor larger than that of the lightest scalar, 
and hence poses no additional cosmological problems. It contributes 
to PAMELA/ATIC signals and therefore dilutes the proportion of these
accounted for by dark photon decays.

\subsection{\it Decays into the SM: collider signatures}
Now let us turn to the possible collider signatures of this kind of model.
Since our DM is not charged under the SM and is heavy, 
it is unlikely that any decay chain at the LHC will end up in the HDM. 
Thus the path at colliders to the dark sector is via the 
lightest MSSM R-odd particle, likely a neutralino.\footnote{Since the
lightest MSSM R-odd particle is unstable in this model, it can also 
be charged or colored. 
Decays of such a particle will therefore result in 
an extra charged or  colored SM state beyond the possible lepton jet.} 
This neutralino will
cascade decay via the supersymmetric gauge kinetic mixing to the LDM
particles, typically including dark-stable particles.  If the
dark photon is among these particles, it will promptly decay back to 
the SM, including the striking ``lepton jet'' signals predicted 
in~\cite{ArkaniHamed:2008qp}\footnote{See~\cite{Bai:2009it} for discussion
of possibility measuring the ``lepton-jets" in colliders}.

As discussed above, in our model the dark photon might also not be 
dark-stable. In that case the dark-stable particle decays back to the 
SM will be dominated by the lightest dark scalar. 
Note that
 the non-renormalizable coupling $\eqref{yukawamix}$ that mediates such 
decays in our model is capable of a wide range of decay lengths. For 
different values of $\Lambda$, they 
are consistent with lepton jets, lepton jets with displaced vertices, 
and decay entirely outside the detector, while still being safe from BBN 
constraints.  

The flavor structure of the couplings $\eqref{yukawamix}$ is critical to 
understanding of which of these possibilities is more likely. The reason is 
that after electroweak symmetry breaking, $S$ exchange can mediate 
flavor-changing neutral current (FCNC) processes, which are very strongly 
constrained by experiments.  If we assume only a rough proportionality 
to SM Yukawa couplings in~\eqref{yukawamix}, as we have mostly assumed earlier,
then adequate suppression of FCNCs from $S$-exchange implies an 
$S$ decay length of at least a meter. If this bound is 
saturated, then many decays of the lightest R-odd sparticle in the 
MSSM will end in lepton signals within the detector. Of course this is 
the edge of the viable parameter space, and the decay length could be 
considerably larger, in which case this source of lepton-jets is not 
realized within the detector. 

If the $S$ couplings in~\eqref{yukawamix} 
are completely anarchic (comparable couplings for 
all pairs of SM flavors compatible with gauge invariance), 
then the $S$ decays may still be exclusively to leptons if $m_S < 
2 m_{\pi}$, as required to avoid HESS bounds on photons from 
HDM annihilations. The couplings can be weak enough to satisfy 
FCNC bounds in hadrons as well as lepton-flavor violation 
constraints. Nevertheless, $S$'s produced via MSSM sparticle 
decays can decay to striking $\mu e$ pairs! 

If for some reason there is 
precise proportionality to SM 
Yukawa couplings (Minimal Flavor Violation) in the $S$ couplings 
of~\eqref{yukawamix}, then FCNC constraints are greatly 
weakened~\cite{DAmbrosio:2002ex},  
in which case 
prompt scalar decays to leptons are robustly possible at colliders. 

Apart from lepton-jet signals, the weak dark-visible couplings can also be 
probed indirectly by their contributions to rare low-energy 
processes~\cite{ArkaniHamed:2008qp,Pospelov:2008zw}. 

\subsection{\it Dark Matter abundance}

Of course, any viable DM model must account for its measured 
cosmological abundance. Fitting this does not overconstrain 
our model, or alter the expected range of phenomenology, 
as we explain here. Note that the dark gauge coupling 
$g_D$ dominates the  annihilations of HDM. Thus with all 
other parameters of our model fixed, we can still fit the 
DM abundance by tuning $g_D$. We have not studied this in 
detail, but by the usual arguments in favor 
of WIMP DM, this tuned value will be $g_D \apprle 1$. 

We have presented our dark spectrum plots for only 
certain $g_D$ values, but these spectra can be realized 
for whatever the fitted value of $g_D$ is. Changes in 
$g_D$ can be offset by the other parameters that feed into 
the scales of light dark matter, namely $\mtridva, \epsilon, 
\eta_i, \lambda, \kappa$. The masses and splittings in the 
HDM sector are independently controlled by $c_i$ and $\mu_A$ 
(see~\eqref{nonreni},~\eqref{amix},~\eqref{ma}) respectively.

The DAMA signal, presumed to be accounted for by the iDM 
mechanism, depends on $\epsilon, m_{\gamma_D}$ for the 
scattering amplitude with ordinary matter, and the 
DM abundance and distribution for its magnitude. 
Since we are considering co-existing DM, where the iDM species 
is just one 
component, the relative abundance of this species can be varied 
by varying its mass ($c_i$-coefficient). This gives us one free 
parameter to fit the strength of the DAMA signal.

\section{Conclusions and Outlook}

In this paper we introduced a framework which can accommodate the 
picture of the dark sector  suggested in~\cite{ArkaniHamed:2008qn}. If the 
dark sector is sequestered from the SUSY-breaking hidden sector, one can easily give an electroweak scale mass to the DM, using the generalized Giudice-Masiero mechanism, and the scale $\frac{\mtridva}{16\pi^2}\sim {\rm GeV}$ naturally emerges for the dark gauge boson. 

We took advantage of this framework to build an explicit model of the dark sector. We checked the viability of the model, verifying compatibility with the cosmological and astrophysical bounds. We emphasize that generic models of the dark sector contain additional stable particles and long-lived particles. Checking that the first is not overproduced while the second does not violate the predictions of BBN is necessary and can only be performed  within concrete models.    

Though our model has been built closely following the lines of the original proposal of~\cite{ArkaniHamed:2008qp}, 
it possesses some distinct features. First, in our model the
 particle which is responsible for the Sommerfeld enhancement is not the particle whose decays dominate the PAMELA signal. We stress that the decaying particle can naturally be significantly
 lighter than the dark photon, successfully evading stringent bounds
 imposed by gamma-rays experiments. Second,
 the kinetic mixing between the visible and dark photons in our construction
is not sufficient in order to produce a viable PAMELA signal, and additional couplings are identified. Such couplings may 
be a generic feature of such models.
We also note that these additional couplings 
between sectors may or may not preserve the lepton jet signature at colliders, 
with little constraint from purely theoretical considerations.

We have focused here in demonstrating that our models are compatible with the 
various data, as we understand them currently. We have tried to do this 
using robust and plausible SUSY dynamics, in an easily 
 generalizable and modular fashion. 
It would be interesting to conduct more precise numerical analysis of our 
models, and to precisely determine experimentally preferred regions in the parameter space,
 and to think through other possible experimental tests of models in 
this class.
It would also be interesting to generalize our constructions, say to
non-Abelian dark gauge group.
 
\hspace{10mm}

{\large \bf Acknowledgments.} We are grateful to Clifford Cheung, Joshua 
Ruderman,  Lian-Tao Wang, and Itay Yavin for bringing
to our attention their work~\cite{Princetongroup}. 
We are also indebted to Kaustubh Agashe, 
Nima Arkani-Hamed, Steve Blanchet, David E. Kaplan, Zohar Komargodski,
Shmuel Nussinov,  
Takemichi Okui, Maxim Pospelov,
Yael Shadmi and Neal Weiner for useful discussions.
A.K. is partially supported by NSF under grant PHY-0801323. 
R.S. is supported by NSF under grant PHY-0401513 and by the Johns Hopkins
Theoretical Interdisciplinary Physics and Astrophysics Center. R.S. is 
also grateful to the University of Maryland Center for Fundamental Physics for hospitality during the research and preparation of this paper.

\bibliography{dark}{}
\bibliographystyle{utphys}

\end{document}